\documentstyle[12pt,epsfig]{article}
\newcommand{\beq}[1]{\begin{equation}\label{#1}}
\newcommand{\eeq}{\end{equation}}
\newcommand{\beqar}[1]{\begin{eqnarray}\label{#1}}
\newcommand{\eeqar}{\end{eqnarray}}

\newcommand{\nn}{\nonumber}
\newcommand{\D}{\partial}
\newcommand{\el}{{\cal L}}
\newcommand{\A}{{\cal A}}
\newcommand{\Ka}{{\cal K}}

\newcommand{\ga}{\gamma}
\newcommand{\de}{\delta}

\newcommand{\ka}{\kappa}

\newcommand{\PR}{Phys. Rev.\ }

\newcommand{\PL}{Phys. Lett.\ }
 \newcommand{\NP}{Nucl. Phys.\ }
\newcommand{\ZP}{Zeit. f. Phys. C\ }

\begin{document}
\vspace*{-2cm}
%\hfill Version 2 
\hfill NTZ 15/97 Revised Version
\vspace{4cm}
\begin{center}
{\LARGE \bf Gluonic Reggeons\footnote{Supported by Deutsche
Forschungsgemeinschaft KI 623/1  \\ and KBN grant 2 P03B 065 10}}\\[2mm]
\vspace{1cm}
R.~Kirschner$^\dagger$ and  L.~Szymanowski$^{\dagger \#}$
\vspace{1cm}

$^\dagger$Naturwissenschaftlich-Theoretisches Zentrum  \\
und Institut f\"ur Theoretische Physik, Universit\"at Leipzig
\\ 
Augustusplatz 10, D-04109 Leipzig, Germany
\\ \vspace{2em}
$^{\#}$
Soltan Institut for Nuclear Studies, Ho\.za 69, 00-681 Warsaw, Poland
\end{center}

\vspace{1cm}
\noindent{\bf Abstract:}
Contributions from gluon interactions, which are non-leading in high-energy 
semi-hard processes, are studied and represented in terms of reggeon exchanges.
Unlike the leading gluonic reggeon, related to the BFKL pomeron,
the non-leading reggeons are sensitive to 
the spin and transverse momentum distributions
 of scattering partons. There are several gluonic reggeons with poles in 
the vicinity of angular  momentum $j=0$ contributing to the
perturbative Regge asymptotics of QCD. We extend the high-energy 
effective action
including sub-leading terms which describe these reggeons and their 
interactions with scattering quarks and gluons in the multi-Regge
approximation.

\vskip.2in
\begin{center}
{\it Dedicated to the memory of V. N. Gribov}
\end{center}

\vspace*{\fill}
\eject
\newpage

\section{Non-leading asymptotics}

In the semi-hard  region, where the momentum scale controlling the 
coupling is large but much smaller than the scattering energy $\sqrt{s}$, 
the behaviour of 
scattering processes is dominated by reggeized gluon exchange. The
exchange of two reggeized and interacting gluons results in the BFKL
pomeron \cite{BFKL}.

In that region the contributions from quark exchange are suppressed 
at least by one power of $s^{1/2}$ for each exchanged fermion. These
contributions dominate, if flavour quantum numbers in the $t$-channel
exclude the pomeron exchange.

There are contributions from gluon interaction, which are down by one 
or more powers of $s$ compared to the leading gluon exchange related to
the BFKL pomeron. This is evident e.g. from the gluon elastic scattering
tree amplitude. Non-vacuum quantum numbers, odd $C$ and $P$ parity, can
also be transferred by gluons.

Consider the scattering with small momentum transfer 
of a gluon or a quark of high momentum 
 on a source of colour fields. The leading
interaction contributes to the amplitude proportional to the first 
power of the large momentum. It conserves helicity of the high momentum 
quark or gluon and it is not sensitive to the details of the colour 
source as its distribution in the transverse (impact parameter) plane
or to its spin structure. However such details are resolved by the
interactions resulting in a contribution to the amplitude suppressed
 by powers of the large momentum compared relative to the leading one.

Our aim is to investigate the non-leading gluon exchange with contributions 
to the amplitude down by one power of $s$ compared to the leading
behaviour. There are measurable quantities of interest, where the non-leading
gluon exchange is essential and contributes not just as a small correction.
The flavour singlet spin structure function $g_1(x)$ is determined in
the region of small $x$ mainly by gluon exchange, where one of the exchanged
gluons is of the non-leading type. The spin structure function of the
photon $F_3^\ga(x)$ related to the helicity flip amplitude is dominated at
small $x$ by exchange of two subleading gluons.

Describing the high energy behaviour it is convenient to consider the
Mellin transform of the amplitude with respect to $s$. It is essentially
the $t$-channel partial wave and the Mellin variable is the complex
angular momentum $j$. Whereas the leading gluon exchange and the BFKL pomeron
correspond to Regge singularities near $j=1$, the non-leading gluon
exchanges studied here appear as reggeons with poles near $j=0$.

In this letter we show that there are several gluonic reggeons near $j=0$.
One of them transfers positive parity like the leading gluonic reggeon.
Three reggeons transfer negative parity. Moreover there are reggeons
carrying colour states different of the one of the gluon. 
In analogy to the BFKL pomeron the colour singlet $t$-channel exchanges
important for physical processes are built from two or more gluonic reggeons
with pair interaction due to emission and absorption of $s$-channel gluons.
Avoiding all
technical details in this paper we are going to discuss the basic ideas of
the approach and to explain the results.

Our tool for investigating the non-leading gluon exchanges is the high-energy
effective action. It has been introduced originally as a convenient starting 
point for improving the leading $\ln s$ approximation for the leading gluon 
exchange \cite{Lev91}. The terms of this action related to leading gluon and quark
exchanges are known in the approximation corresponding to the
multi-Regge kinematics for $s$-channel intermediate state.
The configuration of multi-Regge kinematics of a multiparticle
intermediate state is the one giving the leading $\ln s$ contribution.
This configuration is characterized by large sub-energies of pairs of
particles (equivalently, large rapidity gaps) and restricted momentum
transfers. 
Now we are going to 
include the terms related to the non-leading gluon exchange. A treatment of
the leading gluon exchange going beyond the multi-Regge approximation is given
in \cite{Lev95}. 

The high-energy effective action is obtained from the QCD action by 
separating the momentum modes of the gluon and quark fields into those
relevant for scattering, for exchange and remaining "heavy modes" \cite{KLS}, 
\cite{RKfer}.
The latter are integrated out approximately. 

An important feature of the QCD, which becomes apparent in the
intermediate steps, is the factorizability in the $t$-channel of the
quartic terms describing effectively the high-energy $2 \to 2$ scattering.
Also on the non-leading level this factorizability holds.
This is essential for identifying the reggeons and the scattering vertices
entering the effective action.

The structure of the fermionic terms in the QCD high-energy effective action
can be read off from the ones in supersymmetric Yang-Mills theory \cite{KLS}.
The supersymmetry subgroup of the latter theory compatible with the chosen 
gauge allows to reconstruct fermionic terms out of gluonic ones.
The scattering vertices as well as reggeons form multiplets of this subgroup.
It is interesting that the leading and some non-leading gluonic reggeons are
in the same multiplet.

Some of the technical steps in our procedure can be justified only in
the framework of perturbation theory, the applicability of which is
restricted to the semi-hard region.
Referring to this perturbative Regge region we can give the inverse 
derivatives appearing in the calculations and in the final result a meaning,
since there the typical longitudinal momenta are not small and small
transverse momenta should not give essential contribution either.
Technically inverse derivatives come in by adopting the light-cone
axial gauge and by integrating over the heavy modes. Thus non-local
interactions are an essential feature of the high-energy effective action.
With improvements the effective action concept will be useful also for a 
non-perturbative treatment of peripheral high-energy scattering.

\section{Separation of modes}

We follow the derivation of the effective action invented for the leading
gluon exchange and quark exchange keeping now all terms of the non-leading
(by one power of $s$) contributions \cite{KLS}. The separation of modes
is the first essential step.

 It is convenient to start from the
Yang-Mills action in the light-cone axial gauge $A_- = 0$,
\beqar{1}
    \el & = & \el^{(2)} + \el^{(3)} + \el^{(4)}  \nn \\
   \el^{(2)} & = & -2 A^{a *}(\D_+\D_- - \D\D^*) A^a \nn \\
  \el^{(3)} & = & - \frac{g}{2} J_-^a \A_+^a - \frac{g}{2} j^a \A'^a \nn \\
   \el^{(4)} & = & \frac{g^2}{8} J_-^a \D_-^{-2} J_-^a - \frac{g^2}{8} j^a j^a
\eeqar
We use light-cone components for the longitudinal part of vectors and
complex numbers for the transverse part \cite{KLS}. 
The space-time derivatives are
normalized such that $\D_+ x_- = \D_- x_+ = \D x = \D^* x^* = 1$. The gluon
field is represented by the transverse gauge potential $A^a$, $A^{a*}$.
It enters the interaction terms (\ref{1}) in the combinations
$\A_+^a = \D_-^{-1} (\D A^a + \D^* A^{a*})$, $\A'^a = i(\D A^a - \D^* A^{a*})$ 
and in the currents
\beq{2}
J_-^a = i(A^* T^a\stackrel{\leftrightarrow}{\D}_- A)\;,\;\;\;\; j^a = (A^* T^a A)
\eeq 
In the following we encounter besides of 
the longitudinal components $J_-^a$ also the transverse components
$J^a$, $J^{a*}$ of the vector current (obtained by replacing $\D_-$ by $\D^*$ and $\D$, 
respectively). We use the abbreviation $(AT^a B) =
-i f^{abc} A^b B^c$ with $f^{abc}$ the structure constants of $SU(N)$.

The notations are chosen such that there is a close relation to the leading
terms of the effective action \cite{KLS}: The expression $\A_+^a$ describes the
leading gluonic reggeon and the current $J_-^a$ determines the leading
scattering vertex. We shall see that some of the non-leading gluonic 
reggeons are described  by the expression $\A'^a$ in terms of the original
gluon field $A^a$. 
From the point of view of momentum representation $\A_+^a$ and $\A'^a$ 
represent
the projections of the transverse gauge potential $A^a(k)$, the first parallel
to the transverse part of its momentum $\ka^\mu$ 
and the second orthogonal to $\ka^\mu$.
The non-leading scattering vertices involve besides
of the current $j^a$ other currents like $J^a$, $J^{a*}$.
Removing the redundant field components in the light-cone gauge is convenient
because now the complex field $A^a$ is directly related to the gluonic degrees
of freedom. However introducing this gauge is a technical step which can
be avoided since the effective action is gauge invariant.

We separate the field into modes $A = A_t + A_s + A_1$.
$A_t$ are the momentum modes typical for exchanged gluons, 
$A_s$ are the modes typical for scattering gluons and $A_1$ are the heavy 
modes, which do not contribute directly to the scattering or exchange
and will be integrated out.

The modes are separated according to the multi-Regge kinematics, i.e.
the momentum configuration of a multi-particle $s$-channel (intermediate)
state ($p_l$, $l=0,1,...,n$) giving the dominant contribution in the
leading $\ln s$ approximation. Decomposing the transferred momenta
$k_l = p_A - \sum_{i=0}^{l-1}p_i$ with respect to the (almost light-like)
momenta of incoming particles $p_A$, $p_B$
\beq{0}
k^\mu = \sqrt{\frac{2}{s}}\left(k_+ p_B^\mu + k_- p_A^\mu\right) 
+ \ka^\mu \;\;,
\eeq
the multi-Regge kinematics is characterized by the conditions
\beqar{0'}
&&k_{+n} \gg .... \gg k_{+1}\;\;,\;\;\; k_{-n} \ll .... \ll k_{-1} \nonumber
\\
&&k_{+l}k_{-l} \ll |\ka_l|^2 \;\;,\;\;\;\; s_l = k_{+\;l-1}k_{-\;l+1} \gg
|\ka_l|^2 \nonumber \\
&&\prod_{l=1}^n s_l = s\prod_{l=2}^n |\ka_l - \ka_{l-1}|^2 \nonumber
\eeqar
Here $\ka_l$ denotes the transverse (with respect to $p_A$, $p_B$) part of
the momentum $k$. It is represented by a 4-vector in (\ref{0}) and in the
following it will be represented by a complex number keeping the same
notation. The longitudinal momenta are strongly ordered.
The subenergies $s_l$ are large compared to the transfered momenta.
The longitudinal contribution to the transferred momenta squared is small.
In loops the main contribution from $s$-channel intermediate particles
arises from the vicinity of the mass shell. Therefore the modes
$A_t$, $A_s$, $A_1$ are characterized by the following conditions
\beqar{0''}
&A_t&\;: \;\;|k_-k_+| \ll |\ka|^2  \nonumber \\
&A_s&\;: \;\;|k_-k_+ - |\ka|^2| \ll |\ka|^2 \nonumber \\
&A_1&\;: \;\;|k_-k_+| \gg |\ka|^2 \;\;.
\eeqar
We introduce the mode separation into the action (\ref{1}) by substituting
$A$ by $A_s + A_t + A_1$. The kinetic term decomposes into three, one for
each of the modes, which follows immediately from momentum conservation.

Consider now the triple term $\el^{(3)}$ of the action (\ref{1}).
Introducing the mode separation results in many terms, the most important of
them for our discussion are those where two of the fields have longitudinal
momenta of the same order and the third one has much larger or much smaller
longitudinal momentum. We denote by  $\el^{(3)}_1$ the terms with one of the
three fields in the exchange mode $A_t$ and two in the scattering or heavy
mode $A_s + A_1$. The field in the $A_t$ mode will be written as the last
factor in each of those terms and decomposed into the expressions $\A^a_+$
and $\A'^a$ appearing in the action (\ref{1}),
$\D A_t = \frac{1}{2}(\D_- \A_{t+} -i\A_t')$. Adopting this convention allows
in the following to omit the subscripts ($t$, $s$, 1) referring to the range
of modes. We use the currents $J^a_-$, $J^a$, $J^{*a}$, $j^a$ introduced
above to express the two field in the modes $A_s + A_1$ in some of these
terms. In this way the considered part of the separated triple term
${\el}^{(3)}_1$ can be written by a straightforward rearrangement of terms
as 
\beqar{3}
{\el}^{(3)}_1 & = & -\frac{g}{2} [ J^a_- - \frac{1}{2}\cdot \frac{\D_-}{\D \D^*
}(\D J^a + \D^* J^{a*}) \nn \\
&-& i\frac{ \D^2_-}{ \D^*}(\frac{1}{\D_-}(\D A + \D^* A^*)T^a A) 
-i \frac{\D^2_-}{\D }(\frac{1}{\D_-}(\D A + \D^* A^*)T^aA^*)]\A^a_+ \nn \\
&-& g [j^a + \frac{i}{4}\frac{1}{ \D \D^*}(\D J^a - \D^* J^{a*}) 
- \frac{1}{2}\frac{\D_-}{ \D^*}(\frac{1}{\D_-}(\D A + \D^* A^*)T^a A) \nn \\
&+& \frac{1}{2}\frac{\D_-}{ \D}(\frac{1}{\D_-}(\D A + \D^* A^*)T^a A^*)] \A'^a
\eeqar
It is understood that the second factor in each term ($\A^a_+$ or $\A'^a$) is
in the $A_t$ modes. \\
In the case of scattering with a large momentum component $k_-$ the term
with $J^a_-$ contributes to order ${\cal O}(k_-)$, the terms with $J^a$,
$J^{a*}$ and $j^a$ to ${\cal O}(k_-^0)$ and the other terms to ${\cal
O}(k_-^{-1})$. In describing scattering with large $k_+$ the ordering goes
in the reverse direction.

\section{Quasi-elastic scattering}

In order to identify the perturbative reggeons we study the contributions
to the $2 \to 2$ high energy gluon scattering at the tree level.
Introducing the mode separation into the quartic term $\el^{(4)}$
of the action (\ref{1}) we see that the term with all four fields in
the scattering modes $A_s$ contributes to the elastic scattering.
We write this contribution where two fields have momenta with large $k_-$
and two with large $k_+$ and both close to the mass-shell,
\beq{4}
\el^{(4)}_{scatt} = \frac{g^2}{4} J_-^a \;\frac{1}{\D_-^{2}}\; J_-^a 
- \frac{g^2}{2}j^a\;
j^a - \frac{g^2}{4}j_D^r \; j_D^r \;\;.
\eeq
This result is obtained directly from $\el^{(4)}$ (\ref{1}) 
by selecting the
relevant terms in the mode decomposition and expressing the two fields with
large $k_-$ in terms of a current appearing at the first factor and the two
with large $k_+$ in terms of the second current in each term.
Adopting the convention that the first factor carries the modes with
large $k_-$ we suppress any subscript referring to the modes.
 This expression applies also in the more general case if the
second current in each term is built from fields with all modes 
$A_t + A_s + A_1$ with large
$k_+$. The
currents $j_D^r$ are not in the adjoint representation 
but in the reducible representation (index $r$) appearing as the
symmetric part in the decomposition of the product of two adjoint
representations, (indices $a, b, c, d, e$)
\beqar{5}
         &&   j_D^r = (A^* D^r A) \;\;, \nn \\
 &&(T^e)_{ab}(T^e)_{cd} + (T^e)_{ac}(T^e)_{bd} = (D^r)_{ad}(D^r)_{cb}
\eeqar
The symmetric representation appears because we represent interaction terms 
without a gluon in the $t$-channel in form of $t$-channel exchange.

We write down all quartic terms contributing to the scattering of a gluon
with large $k_-$ with the two other fields in all modes (see Fig.~1)
\beq{6}
{\el}^{(4)}_{tot}  = {\el}^{(4)}_{scatt} + <{\el}^{(3)}_1 \;{\el}^{(3)}_1
>_{A_t} + <{\el}^{(3)}_1 \;{\el}^{(3)}_1 >_{A_1} \;\;.
\eeq
The second term is obtained by substituting the product of the two fields in
the exchange modes $A_t$ by their propagator. The analogous contraction is
done in the last term with $A_1$ modes, where ${\el}^{(3)}_1$ is restricted
to the first order in $A_1$ \cite{KLS}. 
This is the simplest but essential contribution
from the integration over "heavy" modes. In our approximation
we have to sum only the contributions from one-particle intermediate states
out of the modes $A_1$. Technically this can be done by eliminating
$A_1$ using the equations of motion linearized with respect to $A_1$.

We avoid to write here the resulting explicit form of ${\el}^{(4)}_{tot}$
(\ref{6}). We disregard terms ${\cal O}(k_-^{-1})$ and use the condition
that two fields carrying momenta with large $k_-$ are close to mass-shell.
Those two fields we write as the first factor in each term. It turns out
that this factor can be expressed in terms of one of the currents
$J_-^a$, $J^a$, $J^{*a}$, $j^a$, $j_D^r$ in each term.
This observation is a result of an involved calculation. It is an
important step because it is the first signal of the expected
factorizability which we are going to discuss now.

Consider the contribution of ${\el}^{(4)}_{tot}$  to high-energy
scattering where the two other fields are in the scattering modes $A_s$ too,
but with large $k_+$. We obtain from the explicit form of (\ref{6}) with
factorized currents by restricting the two fields carrying large $k_+$
also to the modes $A_s$
\beqar{7}
&&\el^{(4)}_{tot,scatt} 
= \frac{g^2}{8} J_-^a \;\frac{1}{\D \D^*}\; J_{+R}^a -
\frac{g^2}{16} (\D_+ J_-^a) \;\frac{1}{\D \D^*}\; (\D_- J_{+R}^a) \nn \\
&& + g^2 j_s^a \;j_{sR}^a - \frac{g^2}{2}j^a\; j_R^a - \frac{g^2}{4} j_D^r
\; j_{DR}^r + \mbox{{\cal O}}(s^{-1}) \;\;.
\eeqar
The first current in each term carries the scattering modes with large $k_-$
and the second current the scattering modes with large $k_+$. The current
with subscript $R$ is obtained from the expression of the corresponding
current without this subscript by substituting $A^a \rightarrow A^a_R = -
\frac{\D^*}{\D} A^{a*}$. This corresponds to the gauge transformation
leading from the gauge $A_-^a = 0$ to the gauge $A_+^a = 0$.

It is remarkable and essential for the approach that, up to the non-leading
terms, the two fields carrying large $k_+$ can be written in terms
of currents too. The currents of the scattering modes with large $k_+$
carrying the subscript $R$ arise in our calculation as the sum of many terms
transformed by applying the equations of motion for the scattering
particles, i.e. the mass-shell condition.

As the further technical remark it is useful to notice
that the analogous result to Eq. (\ref{7}) without the
substitution $A^a \rightarrow A_R^a$ in the currents with label $R$ is
obtained from the first two terms of (\ref{6}) only if ${\el}^{(3)}_1$ is
truncated by omitting those terms which are ${\cal O}(k_-^{-1})$ in the
large $k_-$ scattering regime.

The other contributions, in particular the ones from the heavy mode
integration, result merely in the gauge transformation, i.e. in the
replacement $A^a \to A^a_R$.

The result
(\ref{7}) has the form expected from parity symmetry: Formulating
the scattering of gluon with large $k_-$ in the gauge $A^a_-=0$ or one
with large $k_+$ in the gauge $A^a_+=0$ should be the same up to
exchanging $+ \leftrightarrow -$ accompanied by the corresponding gauge
transformation. The result exhibits the important feature of $t$-channel
factorizability which is essential for the effective action.

Each of the current times current terms in (\ref{7}) can be viewed as the
result of an elementary $t$-channel exchange (pre-reggeon). Here we can read
off which reggeons arise. The interaction of the pre-reggeons and in
particular their reggeization has to be derived from the further analysis
(Sec. 5).

The first term in Eq.(\ref{7}) represents the leading ${\cal O}(s^1)$
contribution to the quasi-elastic scattering.
Its first factor arises directly from the first term in ${\cal L}^{(3)}_1$
(\ref{3}).
The terms of interest are
the following which contribute as ${\cal O}(s^0)$ to the amplitude.
The second term in Eq.(\ref{7}) is closely related to the leading term, 
both describe positive parity exchange.
Its first factor arises from the second term in ${\cal L}^{(3)}_1$ (\ref{3})
by applying the mass-shell condition.
In the pure gluodynamics considered now 
there are two terms describing colour octet odd parity
exchange (the fourth and fifth term in Eq.(\ref{7})), where
\beq{8}
j_s^a = j^a + \frac{i}{4}\frac{1}{\D \D^*}(\D J^a - \D^* J^{a*}) \;\;.
\eeq
$j_s$ appears in ${\cal L}^{(3)}_1$ (\ref{3}) as the leading terms
involving $\A'$ exchange.
The last term written explicitely in Eq.(\ref{7}) corresponds to the
symmetric gauge group representation in the $t$-channel (compare
(\ref{5}).

Introducing 5 pairs of fields for the pre-reggeon exchanges corresponding to
each term in Eq.(\ref{7})
the quasi-elastic high-energy scattering can be described by the following
effective action
\beqar{9}
&&{\el}_{eff, scatt} = {\el}_{kin.} + {\el}_{s -} + {\el}_{s +} \nn \\
&&{\el}_{kin.} = - 2 A_s^{a*}(\D_- \D_+ - \D \D^* )A_s^a
-2 \A_+^a \D \D^* \A_- + \A_{(+)}^a \A_{(-)}^a \nn \\
&& - \A'^{a(+)}_s \A'^{a(-)}_s + 2\A'^{a(+)}_2 \A'^{a(-)}_2
 + B^{r(+)}B^{r(-)}  \\
&& {\el}_{s -} = - \frac{g}{2}J_-^a \A_+^a - \frac{g}{4} (\frac{\D_+}{\D
\D^*}J_-^a) \A_{(+)}^a -g j_s^a \A_s'^{a(+)} - gj^a\A'^{a(+)}_2 -
\frac{g}{2}j_D^r B^{r(+)} \nonumber \;\;.
\eeqar
${\el}_{s +}$ is obtained from ${\el}_{s -}$ by replacing the labels $+
\leftrightarrow -$ and the currents by their partners with label $R$.
The result for the effective action for quasi-elastic scattering
is checked by observing that each term in (\ref{7}) is reproduced from
(\ref{9}) by integrating out the corresponding pre-reggeon.

Each
pre-reggeon is represented by a pair of fields. The distinction by labels +,
- is related to the fact that the reggeon exchange is oriented in rapidity.
The fields with label + couple to scattering particles with large $k_-$ only
and vice versa.

\section{Including fermions}

In the kinematics of high-energy scattering it is natural to decompose Dirac
fields $\psi$ describing quarks into light-cone parts, $\psi = \psi_- +
\psi_+$, $\ga_- \psi_+ = \ga_+ \psi_- = 0$. Adopting the gauge $A_- = 0$ we
eliminate $\psi_+$ and decompose $\psi_-$ with respect to a spinor basis
$u_{ij}$, $i,j = \pm$ \cite{KLS},
\beq{10}
\psi_- = fu_{--} + \bar{f}u_{-+}\;,\;\;\; \ga u_{-+} = \ga^* u_{--} = 0
\eeq
with two complex component field $f$ and $\bar{f}$. These two fields
describe the two chiralities of a quark.

To describe the essential features of the fermionic terms in the 
high-energy effective action it is enough to look at the case of
supersymmetric Yang-Mills theory, i.e. restricting to one fermion chirality
$f$ and changing to the adjoint gauge group representation.

There is a supersymmetry subgroup (with generators $\de_i$, $\de_r$) 
leaving $A_-$ unchanged and acting on $A$ and $f$ as
\beqar{11}
&&\de_i A = f, \;\;\;\de_i A^* = f^*, \;\;\;\de_i f= 2i\D_- A, \;\;\;
\de_i f^* = 2i\D_- A^* \nn \\
&&\de_r A = f, \;\;\;\de_r A^* = -f^*, \;\;\;\de_r f= -2i\D_- A, \;\;\;
\de_r f^* = 2i\D_- A^* \;\;.
\eeqar
The transformations $\de_i$ and $\de_r$ are ones considered in \cite{KLS}
for the pure imaginary and real transformation parameter, respectively.
The subsequent application of two transformations of the same type results
in the action of the derivative $\D_-$.
This subgroup can be used to reconstruct the fermionic terms in the
light-cone action (\ref{1}) or in the separated triple terms (\ref {3}) from
the gluonic ones. In particular we have used the supersymmetry (\ref{11})
to obtain the fermionic contributions to the currents which we are going to
explain now.

Here we shall not discuss the terms in the effective action related to
fermion exchange con\-si\-de\-red earlier \cite{KLS} and rather concentrate on the
fermionic contributions to the currents appearing in the scattering with
gluon exchange.

By studying the action of the supersymmetry transformations (\ref{11})
to the known gluonic parts of the currents we obtain the fermionic parts
from the condition that applying the supersymmetry transformation twice
must result in the action by the derivative $\D_-$. The fermionic
contributions to the currents ($J \rightarrow J + J_F$, where
$J$ is any of the above gluonic currents) are given by
\beq{12}
J_{-F}^a = (f^*T^af)\;,\;\;\;J_F^{a*} = -\frac{1}{2}(\frac{1}{\D_-} f^*T^a
\stackrel{\leftrightarrow}{\D} f) \;\;.
\eeq
It turns out that with fermions included the current $j^a$ splits up into
two $j^a$ and $\Ka^a$, both with the gluonic part $(A^*T^aA)$ and with the
fermionic parts
\beq{13}
j_F^a = 0\;,\;\;\;\; \Ka_F^a = -\frac{i}{4}(f^*T^a
\stackrel{\leftrightarrow}{\D_-^{-1}} f)\;\;.
\eeq
After inclusion of fermions
the triple terms with gluonic exchange given by Eq. (\ref{3}) still
have the same form (\ref{3})
with the fermionic parts (\ref{12}), (\ref{13}) included in the currents 
and with the replacement
\beq{14}
   j^a \rightarrow \frac{1}{2} (j^a + \Ka^a)\,\,.
\eeq
This splitting of the current $j^a$  
is the reflection of the fact that $j^a$ and $\Ka^a$
belong to the different susy multiplets of currents (see below).

The high-energy scattering terms (\ref{7}) change also by including the
fermionic parts in the currents and by replacing the 4-th term on the r.h.s.
of (\ref{7})
\beq{15}
   j^a \;j_R^a \rightarrow \frac{1}{2}j^a \;j_R^a + \frac{1}{2}\Ka^a \;\Ka_R^a
\;\;\;.
\eeq
According to (\ref{14}) the combination of currents $j_s$ is now
\beq{14'}
j_s^a = \frac{1}{2}j^a + \frac{1}{2}\Ka^a + \frac{i}{4}\frac{1}{\D\D^*}
(\D J^a - \D^* J^{a*})
\eeq
This modified form of ${\el}_{eff, scatt}$ replacing (\ref{7}) in the case of
fermions included is the result of an extensive calculations analogous to the
one leading from Eq.(\ref{6}) to Eq.(\ref{7}) above.

The currents with label $R$ are expressed analogously to the ones without
this label where the fields are replaced by $A_R^a, f_R^a$ \cite{KLS},
\beq{16}
A_R^a = - \frac{\D^*}{\D} A^{a*}\;,\;\;\;\;f_R^a =
\frac{\D^*}{\D_-}f^{a*}\;\;.
\eeq

It is remarkable that the exchange fields $\A_+^a$ and $\A'^a$ together
with two combinations of the exchange modes $f_{t}^a$ form a quartet
under (\ref{11}). Correspondingly, the currents $J_-^a$ and $j^a$ are
the even members of a supersymmetry quartet. $J^{a*}$ is part of a doublet
and $\Ka^a$ is a member of a quartet again. The other even member of
the latter multiplet is $k^a = (\frac{1}{\D_-}f^* T^a \frac{1}{\D_-}f)$
 and it is related to the subasymptotics ${\cal O}(s^{-1})$.
In this way we understand that there are four pre-reggeons in the
gauge group states of the gluon on the non-leading level ${\cal O}(s^{0})$
because there are four different susy multiplets where corresponding
currents $(j^a, J^a, J^{a*}, \Ka^a)$ of this level enter.

 Thus we have obtained that with fermions there are three terms
 contri\-bu\-ting to the high-energy scattering in the colour octet  odd
 parity channel and we find three reggeons near $j=0$ in this channel
  related to the exchange of $\A'$.

  A similar effect of removing degeneracy as in (\ref{14}), (\ref{15})
  leading to one additional reggeon compared to gluodynamics arises in the case
  of the other gauge group representations $j_D^r$.

\section{Reggeon interaction}

Besides of the effective vertices of scattering $\el_{s\pm}$ (\ref{9})
the high-energy effective action involves vertices $\el_p$ of emission of
particles (described by the scattering modes $A_s$) from the pre-reggeons
(related to the exchange modes $\A_t$). These imply reggeization and
reggeon interactions. The final form of the effective action is written
as the sum of terms
\beq{17}
  \el_{eff} = \el_{kin} + \el_{s+} + \el_{s-} +\el_p
\eeq
where $\el_{kin}$, $\el_{s-}$ are given by Eq. (\ref{9}).

A way to obtain $\el_p$ in the multi-Regge approximation is to consider the
inelastic $2 \rightarrow 3$ scattering where the final state particles are
 separated by large rapidity gaps. We construct the terms of order 5 in the
 fields which give an effective description of this inelastic scattering
 $2 \to 3$ on the tree level (Fig.~2)
\beqar{18}
&& {\el}^{(5)} = < {\el}^{(4)}_{tot} \;{\el}^{(3)}_1 >_{A_t} +
    < {\el}^{(4)}_{tot} \;{\el}^{(3)}_1 >_{A_1}   \nn \\
 &&   + < {\el}^{(3)}_1 \;{\el}^{(4)}_{scatt} >_{A_t}
    + < {\el}^{(3)}_1 \;{\el}^{(4)}_{scatt} >_{A_1} \;\;.
\eeqar
Substituting ${\cal L}^{(4)}_{tot}$ (\ref{6}) and ${\cal L}^{(3)}_1$
(\ref{3}) into the formula (\ref{18}) 
results in a lengthy expression which we suppress here.
The technical remark presented in  Sec. 3 and
 referring to Eq.(\ref{7}) applies also here:
the second term contributes mainly to replacing $A^a \rightarrow A^a_R$
in the currents of large $k_+$ particles. The structure of the result
can be read off from the other 3 terms only.

The production vertices $\el_p$ are read off from the result by
representing it as
\beq{19}
  {\el}^{(5)} = < \el_{s-} \;\el_p \;\el_{s+} >_{A_t}
  \eeq
and factorizing the scattering vertices $\el_{s-}$, $\el_{s+}$ in view
of the kinetic terms in the effective action of quasi elastic scattering 
 (\ref{9}), (compare Fig.~2). By extensive calculation we have checked
 that the above factorization property (\ref{19}) holds and have extracted
 the vertices ${\cal L}_p$.

There is a number of terms in $\el_p$ differing by the kind of pre-reggeons
in the two $t$-channels. Here we present as an example the part
referring to the odd parity reggeons ($\A'_s$, $\A_2'$) in both channels
(without fermions) 
\beqar{20}
 \el_p \mid_{\A'_s \A'_2} = - \frac{ig}{4}\{ &&2(\frac{1}{\D}\A'^{(-)}_s T^a \D
\A'^{(+)}_s) + 2 (\D^*\A'^{(-)}_s T^a \frac{1}{\D^*}\A'^{(+)}_s) \nn \\
&+& (\frac{\D^*}{\D}\A'^{(-)}_s T^a \frac{\D}{\D^*}\A'^{(+)}_s) 
+ (\A'^{(-)}_s T^a \A'^{(+)}_s) \nn \\
&+& 4(\A'^{(-)}_2T^a \A'^{(+)}_2)  \\
&+& 2 (\frac{1}{\D}\A'^{(-)}_s T^a \D \A'^{(+)}_2)
   + 2 (\A'^{(-)}_sT^a \A'^{(+)}_2) \nn \\
 &+& 2 (\D^*\A'^{(-)}_2 T^a \frac{1}{\D^*}\A'^{(+)}_s)
+2 (\A'^{(-)}_2T^a \A'^{(+)}_s) \} \frac{1}{\D}A^{a*} + c.c. \nn
\eeqar
The details about the effective production vertices related to
the non-leading ${\cal O}(s^0)$ gluonic reggeons will be given
elsewhere.

\section{Discussion}

At high energy, in the perturbative Regge region, 
gluon interactions contribute not only to the leading asymptotics (Regge
singularity near $j=1$, resulting in particular into BFKL pomeron) but
also to subleading terms in the asymptotic expansion.

By the leading scattering vertex $J^a_-$ only the longitudinal momenta of
the scattering partons can be probed. It does not feel the transverse 
momentum and the spin structure. The currents $(J^a, J^{a*}, j^a, j^a_s, 
\Ka^a, j_D^r)$ which determine the coupling of the non-leading reggeons to the 
scattering gluons or quarks are sensitive to the transverse momenta and 
the helicity of the partons.

Extending the high-energy effective action in the multi-Regge
approximation to the sub-leading terms ${\cal O}(s^0)$ we have identified 
the reggeons arising near angular momentum $j=0$. There is a partner of 
the leading gluon exchange with the same parity property. In the odd parity
channel we find two reggeons in the pure Yang-Mills case carrying the adjoint
colour group representation. With fermions included they split off into three.
The occurence of more than one reggeon in one channel is similar to the case
encountered in the reggeization of a scalar interacting with gauge field 
 \cite{scalar}.

Presently we are missing a good physical understanding of the role of the
different reggeons, in particular in the parity-odd channel. They are 
distinguished by their couplings to the scattering partons. 
Therefore the task which should give the answer to this question is to
look for physical processes where the differences encoded in the
currents $(J^a, J^{a*}, j^a, j^a_s,...)$ show up. Also it would be desirable
to look for physical effects of the reggeons in the symmetric
 colour representations. Among them there is a colour singlet reggeon,
which may be of particular interest, because a single reggeon exchange
 of this type can contribute to physical amplitudes. The other
(non-singlet) reggeons appear only in the exchanges of two or more
reggeons building a colour singlet.
The reggeons interact by the exchange of $s$-channel gluons (production
vertices ${\cal L}_p$).

Our results allow to derive two-reggeon interaction kernels in analogy to
the one appearing in the BFKL equation. The analogon of the BFKL
equation for the exchange of one leading and one non-leading reggeon
results in Regge singularities near $j=0$
of physical significance. They may be related, for example, to
the small $x$ behaviour of (combinations of) structure functions.
An interesting task will be to find out whether the remarkable symmetries
 of the BFKL kernel apply also to the non-leading interactions.

With this analysis we did a step towards a field theoretic understanding
of non-leading Regge singularities. The methods and the validity of the
 results are restricted to the perturbative region.

The high-energy effective action can be considered as a new approach to the
classical question of reggeization in field theory. It provides an
alternative to the approach based on amplitude analysis proposed by
Gell-Mann et al \cite{Gell-Mann} and
 developed in \cite{Grisaru}. Comparing we see that the $t$-channel
factorizability of the amplitudes is a crucial point in both approaches.

\vspace*{1cm}
{\Large \bf  Acknowledgements}

\vspace*{.8cm}

L.Sz. would like to acknowledge the warm hospitality extended to him at
University of Leipzig.

%
%
%\vfill
%\eject

\clearpage
%\section*{}

%%%%%%%%%%%%%%%%%%%%%%%%%%%%%%%%%%%%%%%%%%%%%%%%%%%%%%%%%%%%%%%

\hspace*{3cm}

\begin{figure}[htb]
\begin{center}
\epsfig{file=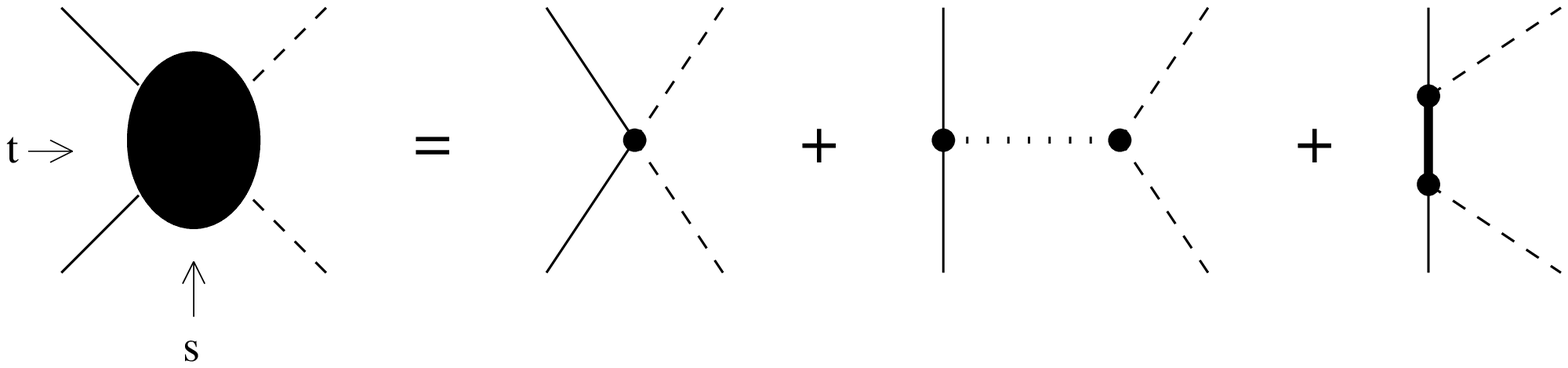,width=15cm}
\end{center}
\vspace*{0.5cm}
\caption{\label{fig-1}
The graphical illustration of Eq. (\ref{6}) for the complete quartic terms.
Different line forms represent different modes: full line - scattered modes,
dotted line - exchange modes, bold line - heavy modes, dashed line - the sum of
all modes.
}
\end{figure}

%%%%%%%%%%%%%%%%%%%%%%%%%%%%%%%%%%%%%%%%%%%%%%%%%%%%%%%%%%%%%%%

\begin{figure}[htb]
\begin{center}
\epsfig{file=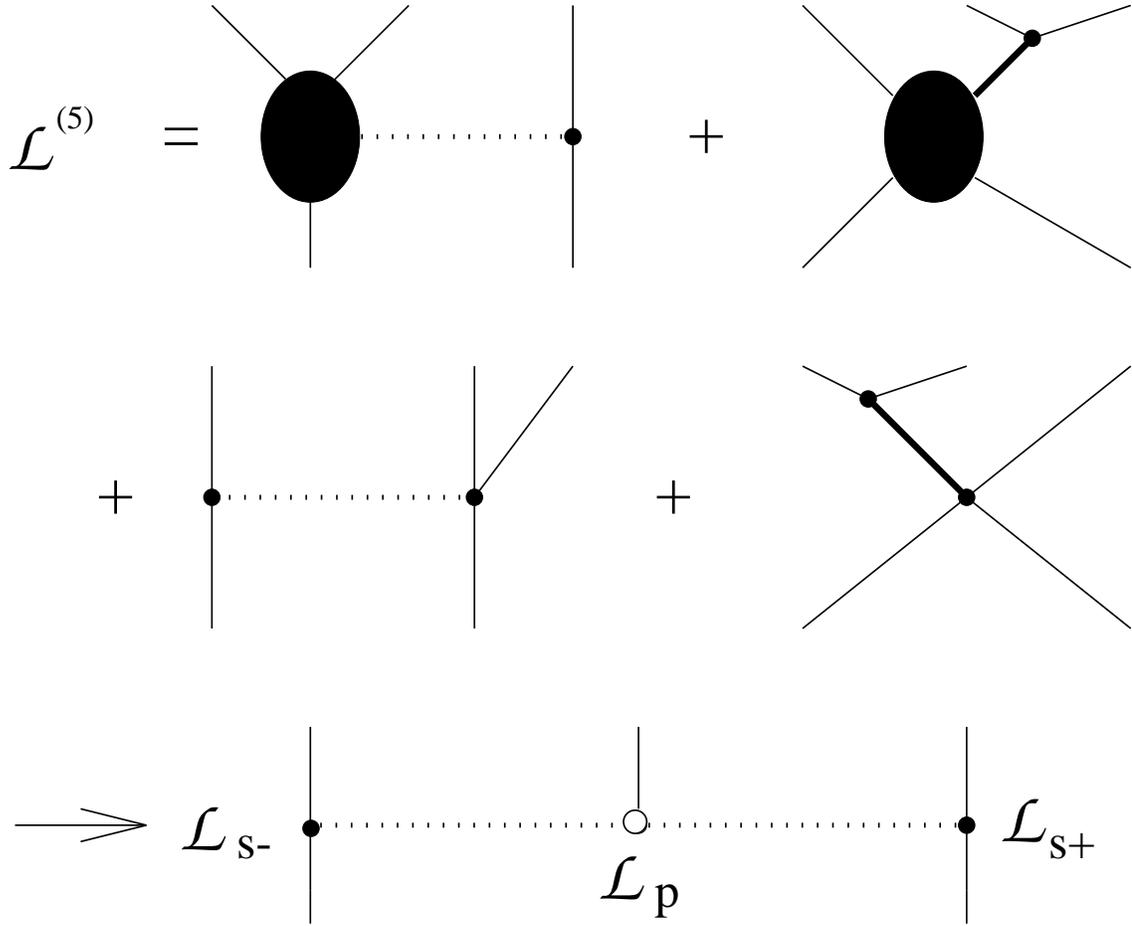,width=15cm}
\end{center}
\vspace*{0.5cm}
\caption{\label{fig-2}
The graphical illustration of the terms describing inelastic scattering 
Eq. (\ref{18}) and
of the factorization leading to the production vertices Eq. (\ref{19}).
}
\end{figure}

%%%%%%%%%%%%%%%%%%%%%%%%%%%%%%%%%%%%%%%%%%%%%%%%%%%%%%%%%%%%%%%

\end{document}